\documentclass[onecolumn]{aastex631}
\usepackage{bbm}
\usepackage{mathrsfs}
\usepackage{amssymb,amsmath}
\usepackage{rotating}
\usepackage{color}

\newcommand\al{\alpha}

\renewcommand\th{\theta}

\newcommand\lam{\lambda}

\newcommand\om{\omega}

\newcommand\De{\Delta}
\newcommand\Th{\Theta}

\newcommand\ie{\emph{i.e.}}
\newcommand\eg{\emph{e.g.}}

\newcommand\beq{\begin{equation}}
\newcommand\eeq{\end{equation}}
\newcommand\bea{\begin{eqnarray}}
\newcommand\eea{\end{eqnarray}}
\newcommand\bal{\begin{align}}
\newcommand\eal{\end{align}}

\newcommand\X{\times}
\newcommand\fr{\frac}

\renewcommand\d{\mathrm{d}}
\newcommand\e{\mathrm{e}}
\newcommand\h{\mathbbm{h}}
\renewcommand\i{\mathrm{i}}

\newcommand\R{\mathbb{R}}
\newcommand\C{\mathbb{C}}
\newcommand\Z{\mathbb{Z}}

\renewcommand\bal{\mbox{\boldmath$\alpha$}}

\begin{document}

\title{Fast radio bursts by stellar wind microlensing of a faint background source}

\begin{abstract}
By assuming the inverse square law of solar wind plasma density as representative of other stars, it is shown that just outside a star the {\it outward} deflection of a passing radio signal at $\nu\approx 1$~GHz (which is capable of penetrating the plasma) is about 5 times larger than the gravitational inward deflection by the star, and the ensuing lens equation which takes both effects into account is a cubic polynomial with three roots and a new strong lensing caustic.  The geometric optics approach is valid for a radio source size $\lesssim 1$~pc.
Microlensing magnification of a steady background source  occurs typically over a timescale of milliseconds, resulting in $\approx 80$ Fast Radio Bursts (FRBs) per day over the whole sky, which can only perturb the isotropy of FRB distribution at the several \% level.  Moreover, repeating FRBs could be triggered by the periodic interception of the line-of-sight of the background source by members of a binary system.  The temporal signatures of such FRBs are consistent with the power spectrum of solar wind density fluctuations on corresponding scales, except the mean density of the wind is a few times higher than the solar value.

\end{abstract}


\author{Richard Lieu}
\affiliation{Department of Physics and Astronomy, University of Alabama, Huntsville, AL 35899}

\section{Introduction}

The deflection of light on astrophysical or cosmological scales has hitherto been considered primarily in the context of gravitational attraction by such mass clumps as stars, galaxies, or clusters of galaxies.  The principal features of this effect are (a) a passing light ray is bent towards the clump; (b) the total deflection angle is twice the Newtonian value due to space curvature which is a General Relativistic concept; (c) the path of the light ray (geodesic) is independent of the light frequency; (d) the deflected ray has a longer time of flight than when the clump was absent; and (e) the ensuing lensing magnification is maximized during source-lens-observer alignment, when the source is not directly visible because a gravitational lens is  generally opaque on the optical axis.
\\

Apart from gravity, light is also deflected on cosmic scales by plasma clumps and voids.  This phenomenon was first investigated in terms of the microlensing by a passing plasma clump with Gaussian profile in electron column density (\cite{cle98}) , and was subsequently applied to the study of Fast Radio bursts (\cite{cor17,er22,ku24}),   and the search for moving dark matter clumps via the gravitational wakes they leave behind, \cite{del24}.  The important difference between the deflection of light by a gravitational field and a plasma lens is that among the five points (a) to (e) listed above, which suit the description of the former process, the exact opposites are true for the latter process, with the exception of (e) which could remain the case for plasma outflow from a star, see below. 
\\

In this paper we consider light deflection by a plasma clump in interstellar space, particularly in respect of radio wave propagation through the immediate environment of stars.  It will be shown, as a conservative estimate of the effect by applying the basic properties of the solar wind to other main sequence stars, that radio waves at frequencies $\nu\gtrsim 1$~GHz from background sources can penetrate the wind but are deflected by it in a frequency dependent way.  
\\


\section{Deflection of light by stellar wind}

In order to gain some general insights on how a passing radio signal is affected by the immediate environs of a star, we begin by using the average steady state properties of the solar wind as representative of stellar outflows.  Indeed, the solar wind density does have a $1/r^2$ density profile at far-field distances (\cite{par58,leb98}), implying an average outflow rate which is constant (but see also other models, namely \cite{new61,sai77,cai09}).   In this paper we adopt the $r\gg 1$ radial profile of Leblanc et al (1998), namely the average solar wind electron number density is assumed to lowest order (\ie~ignoring the influence of the solar magnetic field on the outflow) to be of the form  \beq n_e (r) = \fr{3.3 \times 10^5~{\rm cm}^{-3}}{(r/R_\odot)^2};~r > R_\odot \label{nSW} \eeq  for a fully ionized pure hydrogen plasma outflow, which is generally satisfied by the conditions of a solar wind and corona having $T\approx 10^{5-6}$~K.  

A radio signal skirting a star like the Sun at impact parameter $\De > R_\odot$ is subject to a mean free path against free-free absorption (\cite{kar61}) \beq \lam_{\rm ff} = 5.28 \times 10^{16} \left(\fr{\bar{g} (T, \nu)}{9.82}\right)^{-1} \left(\fr{n_e (r)}{3.3 \times 10^5~{\rm cm}^{-3}}\right)^{-2} \left(\fr{T}{10^6~{\rm K}}\right)^{3/2} \left(\fr{\nu}{{\rm GHz}}\right)^2~{\rm cm},  \label{mfpff} \eeq and Thomson scattering of \beq \lam_{\rm Th} = 4.56 \times 10^{17} \left(\fr{n_e (r)}{3.3 \times 10^5~{\rm cm}^{-3}}\right)^{-1}~{\rm cm},  \label{mfpTh} \eeq  where the default value of the bremsstrahlung Gaunt factor \beq \bar g (T, \nu) = \fr{\sqrt{3}}{\pi} \ln\left(\fr{4kT}{\Gamma h\nu}\right) \label{gaunt} \eeq is evaluated with $\Gamma = 1.781$ at $T=10^6$~K and $\nu = 1$~GHz.  Both mean free paths are far greater than stellar dimensions.  Evidently the radio signal can pass through the plasma in the environment of the star.   

Ignoring relativistic corrections from the bulk and thermal velocities, the plasma frequency is \beq \om_p (r) = \left(\fr{4\pi e^2 n_e (r)}{m_e}\right)^{1/2} = 5.64 \times 10^4 \left(\fr{n_e}{{\rm cm}^{-3}}\right)^{1/2}~{\rm rad~s}^{-1} \label{wp} \eeq 
The refractive index is \beq n(r)= \left(1-\fr{\om_p^2 (r)}{\om^2}\right)^{1/2}. \label{n} \eeq   For a tangential light ray skirting the star at impact parameter $\De > R_\odot$, the deflection angle is \beq \al =  \fr{1}{c} \int \fr{d}{dr} \fr{c}{n(r)} \cos \th ds = 4.17 \times 10^{-5} \left(\fr{\De}{R\odot}\right)^{-2} \left(\fr{\nu}{1~{\rm GHz}}\right)^{-2}~{\rm rad} \approx 8.5~{\rm arcsec}, \label{alpha} \eeq  where $s = \sqrt{r^2 -\De^2}$  is the distance propagated, and $\cos\theta = \De/r$ is the latitude of the ray.  

The expression in (\ref{alpha}) for the deflection angle $\al$ is similar to the one used by Einstein in calculating the bending of light when it passes through a medium of refractive index $n(r)$ (\cite{ein11}) and, as noted by \cite{bla97},  when space curvature can be ignored (which is indeed the case because stellar plasma outflow does not alter the curvature of space, which is due solely to the gravitational potential of the sun).  

The plasma deflection of light as given by (\ref{alpha}) is about 5 times above the amount $4GM_\odot/(c^2 R_\odot) \approx 1.7$~arcsec of gravitational bending of light by the star at its surface.  Moreover, the light is bent {\it away} from the star, whereas gravity does the opposite, because the phase velocity of light in a plasma is $c/n > c$ with the inequality becoming more pronounced towards smaller radii. Thus the conditions responsible for deflection of light are exactly the {\it opposite} of the gravitational deflection of light,  see Figure 1. The combination of these two effects lead to an inward deflection of a passing tangential ray by gravity, and outward deflection by a plasma lens.

When both plasma and gravity are taken into account, the net deflection angle is expressible as \beq \al = \fr{K_1}{\De^2} - \fr{K_2}{\De} \label{al} \eeq where \beq K_1 = \fr{2\pi^2 e^2 n_{\rm e0}}{m_e \om^2} = 2.04 \times 10^{17} \left(\fr{\nu}{1~{\rm GHz}}\right)^{-2}~{\rm cm}^2;~~K_2= \fr{4GM_\odot}{c^2} \approx 6\times 10^5~{\rm cm};\label{K1K2} \eeq with \beq n_{\rm e0} = n_e (r) r^2 =~{\rm constant} \label{ne0} \eeq  by virtue of (\ref{nSW}), and $K_2$ being twice the Schwarzschild radius of the Sun (or a solar mass star).   VLBA observations at 15-43 GHz and in a direction 3 degrees (or $\approx 12~R_\odot$ away from the sun, both being scenarios which minimise the bending of light by solar plasma relative to gravity), were able to confirm the $K_2$ term in (\ref{al}) and (\ref{K1K2}), see (\cite{fom09}).

\section{Microlensing by stellar wind as an origin of FRB}

\begin{figure}
    \centering
    \includegraphics[width=\linewidth]{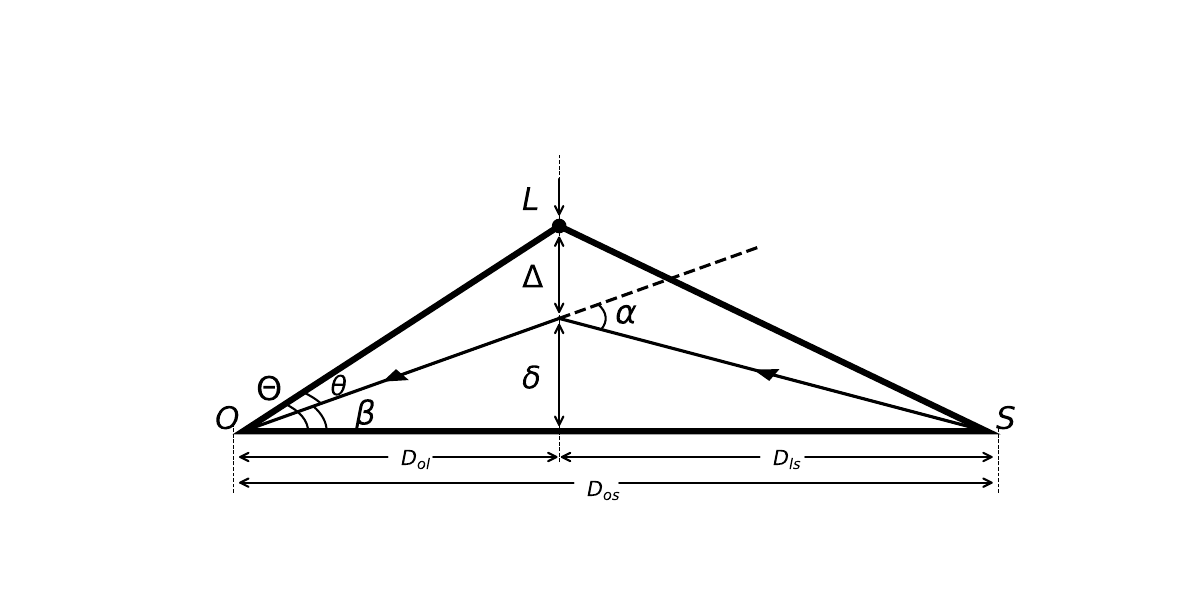}
    \caption{Deflection of light by a plasma lens (effect of gravity is ignored). The locus undertaken by the light signal is marked by arrows.  The shape of it relative to the source and lens positions suggest that the lens deflects light {\it away} from it.}
    \label{triangle}
\end{figure}

In this section we explore the possibility of radio signals from a {\it steady} background radio galaxy being momentarily amplified by the passage across its line-of-sight of a foreground star in the Milky Way to become a FRB, assuming the steady signal traversed the plasma wind and gravitational field of the star, a process known as microlensing.  The idea proposed here differs from \cite{kum24} in a fundamental way: while \cite{kum24} considered foreground plasma lensing to amplify a {\it pre-existing} FRB, our current work suggests that a FRB could also be produced by microlensing signals from a steady (\ie~stationary) background source.  

Let $\th$ be the angular separation between the {\it apparent} direction of a background source and the direction of a stellar plasma lens, and $\Theta$ between the true (unlensed) source directions and the lens, Figure 1.  The impact parameter is then $\De = D_{\rm ol} \th$ where $D_{\rm ol}$ is the distance between the observer and the lens (since the lens is a star inside our Galaxy, its has no redshift, \ie~$z_l$ does not appear in the lens equation),  
which is \beq \fr{(\Th - \th) D_{\rm os}}{D_{\rm ls}} = \fr{K_1}{D_{\rm ol}^2 \th^2} - \fr{K_2}{D_{\rm ol}\th};~{\rm or}~\th^3 - \th^2 \Th  - \fr{K_2 D_{\rm ls}}{D_{\rm ol}{D_{\rm os}}}\th + \fr{K_1 D_{\rm ls}}{D_{\rm ol}^2 D_{\rm os}} =0;~\th \geq 0 \label{lenseq} \eeq  This cubic equation in $\th$ has at most 3 real roots and at least 1, corresponding to a maximum of 3 images and a minimum of 1 for the source.

The crucial task, however, is to find the criterion for maximum magnification (if it exists).  Defining the magnification as the ratio of the flux density of the lensed to unlensed image of the same source, namely $f=F/F_0$, then \beq f= \Big|\fr{\th}{\Th}\fr{d\th}{d\Th}\Big| = \Big|\fr{\th^3}{\Th (3\th^2 - 2\th\Th) -K_2\Th D_{\rm ls}/(D_{\rm ol} D_{\rm os})}\Big|. \label{f} \eeq In general, solution to this cubic would yield at least one image and at most three.  
The magnification of an image becomes infinitely large when the denominator of (\ref{f}) vanishes, \ie~when \beq  3\th^2 - 2\th\Th -\fr{K_2 D_{\rm ls}}{D_{\rm ol} D_{\rm os}} =0. \label{quad} \eeq 
In plasma lensing by a steady stellar wind, $f$ exceeds a certain limit, and each of $\Th$ and $\th$ assumes  {\it one} pair of values, satisfying (\ref{lenseq}) and (\ref{quad}) simultaneously, to mark the `edges' of that limit, as will be shown below.

If the source is much further away than the lens, which is the only scenario being considered in this work, simplication of the two equations is afforded by assuming $D_{\rm os} \approx D_{\rm ls}$.  
To obtain the rate of microlensing of a background radio source, however, one must solve the lens equation (\ref{lenseq}) in conjunction with a magnification lower limit, which we set at $f=200$ for reasons to be explained in the next section.  When (\ref{lenseq}) is used in conjunction with the inequality \beq f>200, \label{f200} \eeq then, many pairs of $\th$ and $\Th$, emerge as solutions.  

With $K_1$ and $K_2$ given by (\ref{K1K2}), there exists a class of solutions\footnote{The only other solution of the lens equation which gives magnification $> 200$ requires an alignment accuracy of source, lens, and observer to a few parts in $10^{23}$, an alignment which can only last 1~ns  assuming that the foreground stellar plasma lens moves across the line-of-sight with peculiar velocity 300 km~s$^{-1}$.  The rapidity of such bursts render them extremely hard to detect, especially since at radio frequencies $\approx 1$ GHz where the coherence time of the data is $\approx 1$~ns, i.e. the intrinsic photon bunching noise barely allows the observer to resolve the signal.  For this reason such a solution was not mentioned in the paper.} where \beq 0 \lesssim |\Th| \lesssim 3.66 \times 10^{-17} = \Th_c; \th\approx 1.135 \times 10^{-10}.  \label{Thth} \eeq The significance of this range of $\Th$ is that a typical foreground star at $D_{\rm ol} \approx 1$~kpc and having an isotropically distributed r.m.s. peculiar velocity \beq u_{300} = 300~{\rm km~s}^{-1}, \label{u300} \eeq   has, under the scenario of isotropic velocity directions, about 25 \% probability of reaching the $\Th$ range of (\ref{Thth}) within one year if it is initially at $\Th\approx 3 \times 10^{-7}$ radians away from the background source.  This is important for the purpose of estimating the rate of microlensing events across the whole sky, see the section 5.  Moreover, at the speed of (\ref{u300}) the time within which $\Th$ satisfies (\ref{Thth}) is $\approx 4$~ms; within this interval, the background source flux is increased by $f \gtrsim 200$ and could manifest itself as an FRB, see below.  Concerning the impact parameter $\De = D_{\rm ol}\th\approx 3.42 \times 10^{11}$~cm, it is just about 5 solar radii away from the center of the star; in fact the caustic is close to the neutral point (\ie~point of no net deflection).

If some FRBs are microlensing events, why have optical identification campaigns failed to find a bright star next to every FRB?   At 1 kpc, the apparent magnitude of the Sun is $m=15$, dimming to $m=20$ at 10 kpc.  Given that most FRBs have a position error $\sim 0.3^{\circ}$, \cite{ami23}, there are too many stars at $m= 15 - 20$ within such an error circle.  A minor subset of FRBs do have good ($< 1^{"}$) positions, see \eg~\cite{nun21}.  Although no optical counterparts were found for these FRBs down to a limiting $m=22$, the sample size is only 8 FRBs, which is too small to  accommodate any claim of inconsistency between theory and data because the theory predicts that $\lesssim 80$ out of several thousands FRBs per day are due to microlensing of faint background radio sources by foreground stars -- see below.

\section{Repeating FRBs and binary microlensing; the role of stellar wind variations}

Although the mechanism proposed here cannot explain repeating FRBs with individual stars (it is also unclear exactly how common such repeaters are among the entire FRB population, \cite{zha23}), the situation improves if one envisages a lone star in a binary system, with a center-of-mass that slowly drifts away (due to orbital motion about the Galactic center) from the line-of-sight of the background radio galaxy on a timescale much longer than the binary period.  Thus \eg~the former could be $\approx 1$ year (but beware the shortest main sequence binary period is of order a fraction of a day, \cite{nor11}), while the latter years because the projected center-of-mass velocity transverse to the line-of-sight can be as small as a few $1$~km~s$^{-1}$.   In this way the two members of the binary system would each be able to pass through the line-of-sight to the background source repeatedly until the whole system drifted away completely.  During a passage, the smooth and steady stellar wind model currently employed would only predict one FRB of millsecond duration, (\ref{Thth}).  Observations point to many bursts, with inter-burst interval of order hours to days, which may be due to time variability in the stellar wind density.  

To be specific, we enlist the repeater FRB 20240209A as case study, because this source has a low dispersion measure commensurate more with an interstellar than intergalactic column density; see \cite{pal25} where the following source properties may also be found.  There are apparently three timescales associated with FRB 20240209A, namely the \beq P_1 \approx 6~{\rm days};~P_2 \approx 30~{\rm days}~P_3 \approx 120~{\rm days}, \label{P123} \eeq where $P_1$ is the average inter-burst interval, $P_2$ is the average window of burst activity with this window returning after the time $P_3$.  If the behavior given by (\ref{P123})  is due to a solar mass binary system in which each star intercepts the line-of-sight of the background radio source every interval of $P_3$ (\ie~$P_3$ is half the binary period), the radius of the (presumably circular) binary orbit would be  $R=0.481$ AU and the orbital velocity $v_b \approx$ 22 km$^{-1}$.  Assuming a distance of $D_{\rm ol} \approx$ 1 kpc to the binary, then, the angular constraint for $\Th$ in (\ref{Thth}) implies a period of $D_{\rm ol} \Th_c/v_b \approx 50$~ms in which the background source is magnified by more than 200 times, and one observes a FRB as a result.  Yet that is not a result of significance here -- there are many more bursts to follow.  

The next and much more important question is whether the average time between consecutive bursts, namely $P_1$ in (\ref{P123}), can be attributed to stellar wind variability.  According to the value of $v_b$ above and the assumption that the source's line-of-sight lies on the binary orbit, in the time $P_1$ from the onset of the interception criterion (\ref{Thth}) 
the star would have moved a distance $P_1 v_b$ further off-axis (\ie~further away from the (undeflected) line-of-sight to the source, with the distance being measured on the lensing plane), which means $|\Th| = P_1 v_b/D_{\rm ol} = 3.85\times 10^{-10}$.  With this new value of $\Th$, (\ref{lenseq}), (\ref{f}), and (\ref{f200}) become unsolvable, \ie~ there is no possibility of attaining an appreciable lensing magnification, {\it unless} the parameter $K_1$, or equivalently the stellar wind plasma density, is increased by $\approx$ 50 times.  More precisely, at time $P_1$ from the last on-axis configuration of $\Th =0$  the solution to (\ref{lenseq}) that gives $f>200$ magnification is \beq \Th = 3.85\times 10^{-10};~8.15\times 10^{-9} \leq \theta \leq 8.29 \times 10^{-9};~49.08 \leq \xi \leq 49.11 \label{P1soln} \eeq  where $\xi$ is the value of $K_1$ in units of the average solar value.  On the other hand, at time $P_2$ from $\Th=0$ the solution is \beq \Th=1.925\times 10^{-9}; 8.80 \times 10^{-9} < \theta < 8.83 \times 10^{-9};~54.079 \leq \xi \leq 54.080. \label{P2soln} \eeq Recall that the $P_{1,2,3}$ values are in (\ref{P123}). 

Given (\ref{P1soln}) and (\ref{P2soln}), for FRB 20240209A one may assume that both stars have their $K_1$ parameter varying between the 7 and 50 times the solar value of (\ref{K1K2}) on a timescale (half period) $\approx 3$ days, corresponding to a mean fluctuation power spectral density (PSD) of $1.25 \times 10^{8}$~cm$^{-3}$~Hz$^{-1}$ at the frequency $\nu = 2 \times 10^{-6}$~Hz and for bandwidth $\delta\nu =0.1\nu$, on par with the extrapolated value of $10^8$~cm$^{-3}$~Hz$^{-1}$ from the solar wind fluctuation PSD in Figure 2 of \cite{che12}.  Concerning the period $P_2$, the fact that no activity exists beyond the $P_2$ time window, apart from the 4 month orbital periodicity of $P_3$ which returns a binary member to the line-of-sight, suggests that the variability subsides downstream of the stellar wind, a phenomenon which has also been observed in the solar wind itself, \cite{col82}.

Three questions remain.  First is the origin of the millisecond FRB fluctuations.  From (\ref{P1soln}) and (\ref{P2soln}), one sees that if the plasma density variation as indicated by the $\xi$ range is $\gtrsim 0.03/50 \approx 6 \times 10^{-4}$, the magnification $f$ will fall below the threshold of burst observability (\ref{f200}).  Thus, if such a fluctuation in the density occurs over the millisecond timescale, a FRB would in principle be induced by microlensing.  This corresponds to a PSD of $4\times 10^{-9}$~cm$^{-3}$~Hz$^{-1}$ at the frequency $\nu = 1$~kHz, which is in agreement with the power law index of $-2.7$ for solar wind PSD at the same frequency, see Figure 2 of \cite{che12}.  

Second remaining question is the center-of-mass velocity of the binary as the system orbits the Galactic center.  The projected velocity along the direction transverse to the line-of-sight must be sufficiently small that within the time $P_3$ as given by (\ref{P123}) the center-of-mass has moved by an amount negligible compared to the binary radius of 0.481 AU (see above).  This practically constrains the transverse velocity to $\lesssim 1$~km$^{-1}$, a small but by no means impossible upper limit which ensures the binary stars can intercept the quasar line-of-sight even after 7 cycles, each lasting the time $P_3$, which is the longest of all 3 recurrence periods of FRB 20240209A.

Third question is why the time variability of the stellar wind did not play any role in accounting for one-off FRBs?  Since the repeating FRBs require a minimum plasma density of $7$ times the solar value (at the same radius downstream of the star) to trigger, lone stars with a wind of solar characteristics are the likely cause of those non-repeating FRBs that may adequately be modeled, at least in terms of their salient features, by a smooth and steady wind.

\section{Upper limit on the source size; narrow band emission}

Is there any limit on the source size for the aforementioned strong lensing mechanism, which is based on geometric optics, to work?   It turns out the large magnification in (\ref{f200}) is valid only if the angular source size of the source is less than the Fresnel diffraction limit $\sqrt{\lambda/D_{\rm ol}}$.  Equivalently, the criterion is that the physical size $d_s$ of the source must satisfy the inequality $d_s < d_s^{\rm max}$ where \beq  d_s^{\rm max} = D_{\rm os} \sqrt{\fr{\lambda}{D_{\rm ol}}}, \label{Fresnel} \eeq see (\cite{gri18,fel23}).
Assuming $D_{\rm os} \approx 10$~Gpc, and
$D_{\rm ol} \approx 1$~kpc, (\ref{Fresnel}) gives \beq d_s^{\rm max} \approx \left(\fr{1}{\nu}\right)^{1/2}~{\rm pc} \label{srcsize} \eeq  as the upper limit on the source size for the magnification to be as large as the geometric optics result of (\ref{f200}) and (\ref{Thth}).  In the case of $D_{\rm os} \approx 10$~Gpc, $D_{\rm ol} \approx 10$~kpc, and $\nu = 1$~GHz, (\ref{Fresnel}) gives $d_s^{\rm max} \approx \sqrt{0.3/(\nu/{\rm GHz})}$~pc as the upper limit on the source size.  According to (\cite{hsu23}), the core size of bright blazars at $15$ GHz frequency is $\approx 5-50$~pc, but because the emphasis of this work is microlensing amplification of faint extragalactic radio sources which are expected to have smaller core emission regions, sizes $\approx 1$~ pc seem possible.  Thus the source could well be small enough for geometric microlensing to take place.

An interesting property of the microlensing radio signal, apart from the millisecond duration which falls smack on the temporal regime of FRBs, is the possibility of a narrow spectral range of the burst.  
Thus, it is not inconceivable that the actual quasar radio emission size $d_s$, while also monotonically decreasing with increasing frequency $d_s$, behaves in such a way that $d_s > \d_s^{\rm max}$ at high {\it and} low frequencies, so that effective microlensing amplification of the signal can only take place within a narrow range of intermediate frequencies.
As case in point, consider the following empirical (and monotonic) relationship between $d_s$ and $\nu$,  \beq d_s =  0.9\sqrt{\nu_9} \left(\fr{1}{\nu_9^2} + 0.7 \fr{\ln \nu_9}{\nu_9} \right), \label{dS} \eeq 
where $\nu_9$ is the frequency  in units of $10^9$~Hz, which yields the correct trend.  Yet at the same time
a comparison between (\ref{srcsize}) and (\ref{dS}), as shown in Figure 2, reveals that $d_s > d_s^{\rm max}$ only within the frequency range $0.8 < \nu < 3$~GHz, \ie~microlensing only occurs within this range of frequency, which could then account for the narrow frequency range of some FRBs as reviewed by \cite{kum24}, although in this instance the plasma will have to be in the interstellar medium (ISM), since under the current scenario the FRB is produced at the distance of a foreground Galactic star which microlenses a faint and non-bursting background radio galaxy.  Alternatively, the phenomenon of narrow band emission could also be explained in terms of the lensing of a coherent FRB burst by a foreground plasma screen, \cite{kum24}, in which case the dispersion measure under this scenario could be higher than ISM levels of a few $\times 100$~cm$^{-3}$~pc, to reach a few $\times 10^3$~cm$^{-3}$~pc if the plasma screen is extragalactic.
\begin{figure};
    \centering
    \includegraphics[width=0.7\linewidth]{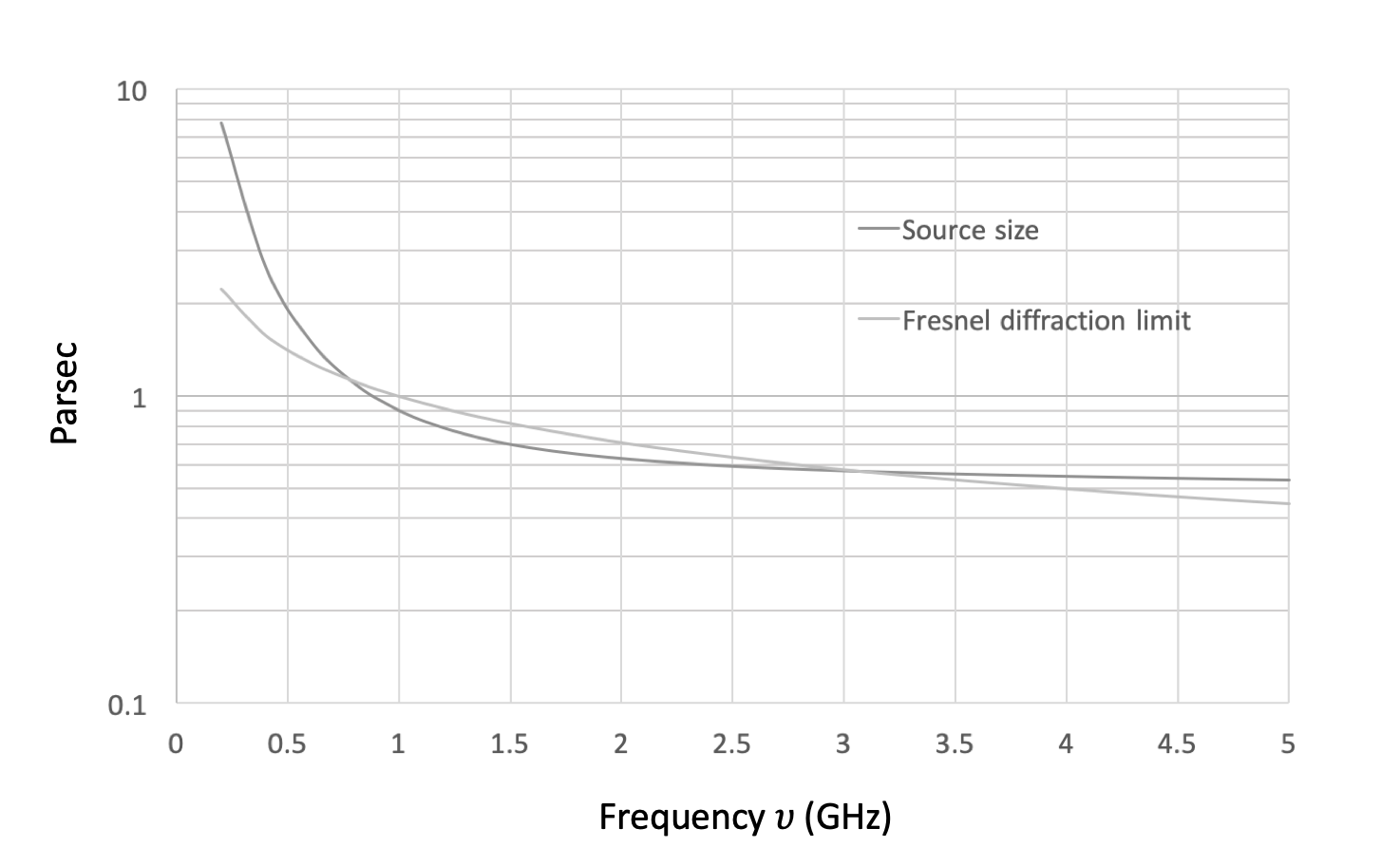}
    \includegraphics[width=0.7\linewidth]{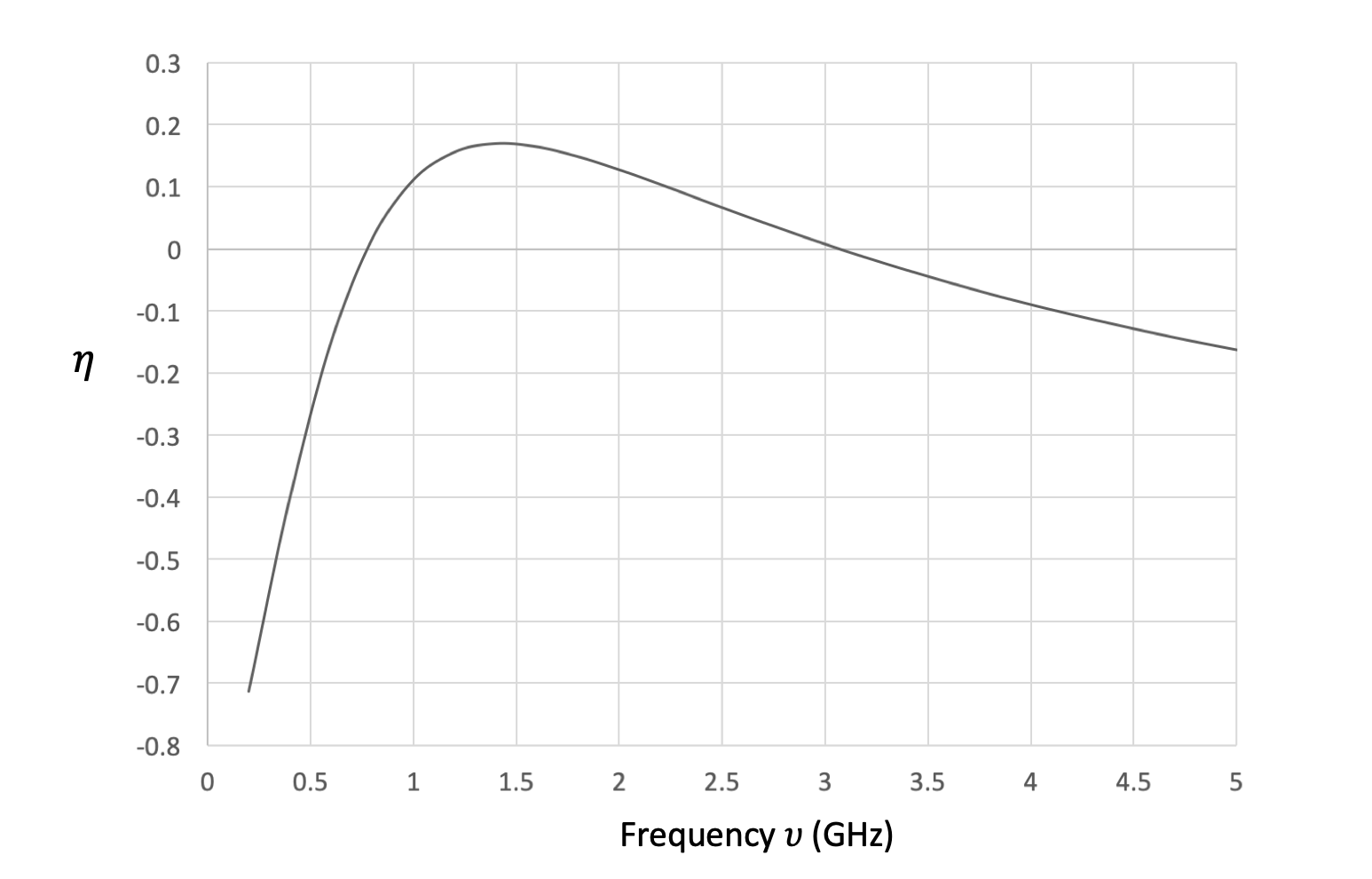}
    \caption{(a) A comparison of the source size $d_s$ against the Fresnel diffraction limited source size $d_s^{\rm max}$ given by (\ref{Fresnel}) at radio frequencies $\nu$ (in GHz).  (b) The dimensionless quantity $\eta = (d_s - d_s^{\rm max})/d_s$ is plotted against $\nu$.
    Within the narrow bandwidth where $\eta > 0$, the entire source fits within the Fresnel limit, and the microlensing magnification $f$ as calculated by the classical (geometric optics) method of this paper is valid.  Outside this bandwith $f$ lies below its classical value, resulting in reduced magnification, rendering the source invisible.}
    \label{source size}
\end{figure}





\section{Microlensing probability}

To estimate the expected rate of FRB occurrences attributable to the microlensing of faint and distant radio galaxies by the wind of foreground stars, we first find the
probability of a radio source appearing positionally coincident with a foreground star, as \beq P= \int n(L_s) dL_s \int_0^{\Th_{\rm max}} \Th d\Th \int_0^{r_{\rm max}} 2\pi r^2 dr. \label{prob} \eeq where $\Th_{\rm max} = 3 \times 10^{-7}$, as explained in the end of the last section, and the luminosity function of the radio loud elliptical galaxies is provided by \cite{dic88},  as 
\beq L_s n(L_s) = 10^{-6} \left(\fr{L_s}{10^{25}~{\rm W~Hz}^{-1}}\right)^{3/2}~{\rm Mpc}^{-3} \label{lumfunc} \eeq for $10^{24.75} < L_s < 10^{26.25}$~W~Hz$^{-1}$.  The comoving distance to the furthest galaxy, $r_{\rm max}$, is simply set to 10 Gpc. 
This approach yields \beq P= 1.48\times 10^{-7} \left(\fr{r_{\rm max}}{10~{
\rm Gpc}}\right)^3 
\label{P} \eeq  The reason for setting the minimum threshold magnification at 200 in (\ref{f}) is to enable the faintest galaxy, with $L_s = 10^{24.75}$~W~Hz$^{-1}$ to be detected at the  brightness sensitivity limit $B=1$~mJy of a large telescope from a distance of $r_{\rm max} = 10$~Gpc via the conversion formula \beq B = \fr{L_s}{4\pi (1+z)^2 r^2} \label{LtoB} \eeq with $1+z \approx 10$ and $r\approx 10$~Gpc.  Evidently, the assumptions here are very conservative, and $P$ is likely a lower limit.

Nevertheless, since any of the $N\approx 2 \times 10^{11}$ stars (\cite{bla16}) with solar characteristics and having a distance $D_{\rm ol} \gtrsim 1$~kpc from the observer can microlens a background source, the quantity $NP$, with $P$ as given in (\ref{P}), is then approximately the number of background ratio sources which will flare as FRBs by stellar wind microlensing within one year\footnote{See the end of the previous section for the explanation of this timescale.}.  Thus the number of FRBs expected to occur in a day across the whole sky, due to foreground stellar wind microlensing,   is \beq \fr{NP}{365} \approx 80. \label{NFRB} \eeq   This is only a few percent the total number of detectable FRBs across the sky is several $\times 10^3$ (\cite{lor24}), meaning that only a small  fraction of FRBs is caused by stellar microlensing.  In any case the isotropy of the distribution of FRBs (\cite{cal17}) is inconsistent with them being due to a foreground Galactic effect, since stars are mainly found on the Galactic disk.

\section{Conclusion}

This paper proposes a conservative limit to the role of stellar wind outflows in the microlensing of {\it steady} background radio sources to account for some of the FRBs, including ones with narrow spectra.   Specifically it is demonstrated that the plasma wind outflow can deliver a brief (of order ms) but very large amplification of the brightness (loudness) of a background radio galaxy by a star which crosses one's line-of-sight at some typical peculiar velocity $\approx 300$~km~s$^{-1}$.  The event rate is estimated to be only $\approx 1 - 2$ \% of the total FRB rate across the sky. Moreover, plasma microlensing can also lead to repeating FRBs if the foreground lens comprises a binary system with a variable stellar wind from both members. The power spectrum of fluctuations appear to be consistent with the observed temporal characteristics of repeating FRBs. 

However, although microlensing could account for those FRBs with narrow spectra and lower than extragalactic dispersion measures (due to the burst being located at stellar distances), the mechanism cannot currently explain why repeating FRBs have narrower spectra than non-repeating ones. 

Plasma microlensing is by no means sufficient to explain all forms of FRB manifestations, yet it is nevertheless a viable mechanism worthy of consideration, since FRBs as a class of sources in its entirety is likely caused by a variety of astrophysical phenomena, \cite{ple21}.




Lastly, concerning the stellar corona, \cite{leb98} shows that the solar corona has a much higher density and steeper density profile at $1~R_\odot$ than the solar wind, which potentially could lead a higher efficacy in microlensing than solar wind.  Nevertheless, observations of {\it stellar} coronae are hard to perform with sufficiently high resolution to enable a firm conclusion that they are all or mainly like that of the Sun (\cite{sch97}); whereas the solar model of stellar wind seems more robust, \cite{sai20}. Consequently, stellar coronae may, or may not, greatly raise the event rate to significantly perturb the isotropy of observed FRBs.  We do not yet know, nor is this the purpose of the current work, which is to point out the role of plasma microlensing in the brightening of some background radio sources on millisecond timescales.   Moreover, if the Sun is any reliable indicator, stellar coronae radically differ from stellar wind in terms of the complexity of it's radial density profile to warrant a separate paper on the subject.


\newpage


\end{document}